\documentclass[12pt]{article}
    \newcommand{\be}[1]{\begin{equation}\label{#1}}
    \newcommand{\ba}[1]{\begin{eqnarray}\label{#1}}

    \newcommand{\rd}{{\rm d}}

    \newcommand{\re}{{\rm e}}
    \newcommand{\pa}[1]{\left(#1\right)}
    \newcommand{\paq}[1]{\left[#1\right]}
    \newcommand{\pag}[1]{\left\{#1\right\}}
    \newcommand{\av}[1]{\langle#1\rangle}
    
    \newcommand{\m}{{\rm m_{\rm P}}}

    \def\ee{\end{equation}}
    \def\ea{\end{eqnarray}}

\usepackage{graphicx,latexsym}
\usepackage{amsmath}
\usepackage[T1]{fontenc}
\usepackage[utf8]{inputenc}
\usepackage{authblk}
\usepackage{amssymb}
\usepackage{xcolor}
\begin{document}
\title{Quantum-cosmological corrections to inflationary spectra from Gaussian wave packets}
\author[1]{Sebastien Sonet\thanks{Sebastien.Sonet@studio.unibo.it}}
\author[2]{Alessandro Tronconi\thanks{Alessandro.Tronconi@bo.infn.it}}
\author[2]{Giovanni Venturi\thanks{Giovanni.Venturi@bo.infn.it}}
\affil[1]{Dipartimento di Fisica e Astronomia, Via Irnerio 46, 40126 Bologna,
Italy}
\affil[2]{Dipartimento di Fisica e Astronomia and INFN, Via Irnerio 46, 40126 Bologna,
Italy}

\date{}
\maketitle

\begin{abstract}
In the context of quantum cosmology, we study quantum entanglement in solutions of the Wheeler-DeWitt equation for a homogeneous inflaton-gravity system coupled to perturbations.  The homogeneous degrees of freedom are described by a superposition of Wheeler-DeWitt solutions associated with a Gaussian wave packet, which is entangled with the quantum state of the perturbations and this may lead to potentially observable effects on the inflationary spectra. The expectation value of the resulting state closely follows the classical homogeneous inflaton trajectory. We then restrict our analysis to two simplified models that admit a future de Sitter attractor, while neglecting nonadiabatic effects and contributions of order $\mathcal{O}(\hbar^2)$ in the Mukhanov-Sasaki equation. Analytical expressions for the resulting spectra are obtained by marginalizing over the homogeneous inflaton variable and can be evaluated in regimes where reliable approximations are available. Small deviations from the standard semiclassical spectra are indeed generated, and these are significantly larger than the quantum-gravitational corrections usually considered in this context.
\end{abstract}
\section{Introduction}
The standard description of the very early Universe assumes that small quantum fluctuations of the matter--gravity vacuum evolve on top of a classical, homogeneous background. In the absence of a universally accepted theory of quantum gravity, this hybrid framework has proved remarkably successful. In particular, the presence of a fundamental, or effective, scalar degree of freedom---the inflaton---can drive the primordial epoch of accelerated expansion known as inflation \cite{Starobinsky1980,Guth1981,Linde1982,AlbrechtSteinhardt1982}, give rise to the primordial quantum fluctuations that seed the observed cosmological structures \cite{MukhanovChibisov1981,MukhanovFeldmanBrandenberger1992}, and eventually decay into ordinary matter, thereby reheating the Universe. The resulting primordial spectra accurately account for the observed CMB anisotropies and the large-scale distribution of matter \cite{Planck2018VI,Planck2018X}.

Inflationary cosmology and black hole physics are widely believed to encode signatures of quantum gravity, which is generally expected to become relevant at energies approaching the Planck scale. Moreover, the high degree of symmetry characterizing these systems often allows the adoption of simplified mathematical descriptions. For instance, the early Universe must be extremely homogeneous and isotropic for inflation to occur and successfully reproduce the observed CMB anisotropies. Under these assumptions, the matter--gravity system can be reduced to two homogeneous degrees of freedom together with an infinite set of perturbative modes describing the small fluctuations around the background. Within this minisuperspace framework, the quantum nature of spacetime can be investigated by canonically quantizing the homogeneous degrees of freedom. This procedure leads to the Wheeler-DeWitt (WdW) equation \cite{DeWitt1967}, which has the form of a time-independent Schr\"odinger equation and whose solution is commonly referred to as the wave function of the Universe. Early applications of the Born-Oppenheimer method to homogeneous matter--gravity systems were developed in Refs.~\cite{BroutVenturi1989,Venturi1990,BertoniFinelliVenturi1996}. More recently, several approaches have incorporated cosmological perturbations within the Wheeler-DeWitt framework, typically relying on a Born-Oppenheimer factorization and a semiclassical expansion of the homogeneous wave function~\cite{BornOppenheimer1927,KieferSingh1991,KTV2018BO,KTVV2019Time,KTV2021BO}. These developments have opened the possibility of investigating quantum-gravitational corrections to the inflationary perturbation spectra~\cite{AlberghiCasadioTronconi2006,KTV2013,KTV2014,KTV2015,KTV2016,KieferKramer2012,BiniEtAl2013,BrizuelaKieferKramer2016dS,BrizuelaKieferKramer2016SR,Calcagni2013,ChataignierKramer2021}. These articles predict broadly similar quantum-gravitational corrections, although their detailed form depends on the approximation schemes adopted to simplify the WdW equation\footnote{the quantum-gravitational effects are obtained by factorizing the homogenous wave function and the wave function which describes inhomogeneities which is shown to finally satisfy a modified Mukhanov-Sasaki equation.}. The resulting corrections to the inflationary spectra are found to be proportional to $H^2/\m^2$, where $H$ is the Hubble constant during inflation, multiplied by the $k$-dependent contribution $(\bar k/k)^3$ where $\bar k$ is an unspecified wave number. In particular, Refs.~\cite{KTV2014, KTV2015, KTV2016,ChataignierKramer2021} also find a (time-dependent) logarithmic contribution. The possible origin of this potentially large contribution is discussed in Ref.~\cite{ChataignierKramer2021}, where different approaches to dealing with it are proposed. The reference wave number $\bar k$ can be associated with the pivot scale used in CMB analyses or other large scales. Owing to their inverse-cubic scale dependence, these corrections become negligible at sufficiently large wave numbers, corresponding to higher CMB multipoles, where the standard inflationary predictions agree well with observations. The emerging quantum-gravitational effects are tiny since they are suppressed by the ratio $H^2/\m^2$ which is constrained to be of order $10^{-10}$ or smaller by the observational upper bound on the tensor-to-scalar ratio. Restoring physical units while keeping $c=1$, the dimensionless suppression factor takes the form $\hbar^2 H^2/\m^2$, where $H$ has dimensions of inverse time and $\m$ of energy.

In this paper we take a further step by treating the homogeneous inflaton (or other combination of homogeneous variables) as a quantum degree of freedom described by a Gaussian wave packet, constructed as a superposition of Wheeler–DeWitt solutions. An appropriately localized Gaussian wave packet can reproduce the classical evolution of the homogeneous inflaton in the $\hbar\rightarrow0$ limit, provided its field and momentum dispersions vanish in this limit. Thus, if proper limits are considered, this construction naturally extends both the standard approach to the study of inflationary dynamics, wherein the homogeneous inflaton and scale factor are taken as c-numbers, and the semiclassical treatment of the WdW equation where the homogeneous degrees of freedom are described by the WKB state $\sim \re^{iS}$ while reproducing the standard classical and semiclassical evolution in the appropriate limit. It also allows us to investigate genuine quantum effects associated with the finite width of the wave packet and their impact on the inflationary perturbation spectra. We restrict our analysis to two simplified scenarios in which the cosmological expansion approaches a de Sitter attractor, and we show that the resulting inflationary spectra can be significantly modified depending on the width of the wave packet. More general and realistic models, as well as inflationary backgrounds departing from the de Sitter limit, are left to future work. Let us note that, unlike in the standard classical treatment, the homogeneous inflaton is therefore not associated with definite (classical) field and momentum values, but is itself described by a quantum state whose finite spread leaves observable imprints on the inflationary perturbation spectra. 

The manuscript is organized as follows. In Sec.~II we introduce the formalism and derive the WdW equation for a minimally coupled inflaton, neglecting inhomogeneities. The case of a constant potential is solved exactly, and Gaussian superpositions for the wave function of the Universe are constructed. In Sec.~III we extend the analysis to induced gravity, in particular to a nonminimally coupled inflaton with a quartic potential. In Sec.~IV cosmological perturbations are included and the Mukhanov-Sasaki (MS) equation is derived under suitable approximations. In Sec.~V we discuss the resulting spectrum of the MS variable and compare it with the standard semiclassical prediction. Finally, in Sec.~VI we summarize our results and present our conclusions.

\section{Formalism}
Let us consider the homogeneous action for a minimally coupled inflaton-gravity system
\be{genact}
S=\int d^4 x\sqrt{-g}\left[-\frac{\m^2}{12}R+\frac{1}{2}\partial_\mu\phi\partial^{\mu}\phi-V(\phi)\right]
\ee
where $R$ is the Ricci scalar and we adopt the following Robertson-Walker metric
\be{metric}
ds^2=a(\eta)^2\pa{d\eta^2-\vec {\rd x}\cdot \vec {\rd x}}.
\ee
The following expressions for the momenta are then obtained:
\be{momG}
\pi_a=-\m^2 a',\;\pi_\phi=a^2\phi',
\ee
leading to the Hamiltonian constraint
\be{hamG}
\mathcal{H}=\frac{\pi_\phi^2}{2a^2}-\frac{\pi_a^2}{2\m^2}+a^4 V=0.
\ee
Upon canonically quantizing the system one is led to the WdW equation
\be{wdwG}
\left(\partial_A^2-\partial_F^2+2\m^{\!\!\!-4}\re^{6A}V\right)\Psi(A,F)=0\,,
\ee
where a convenient ordering for the kinetic term associated with the scale factor has been chosen and the new variables $A\equiv\ln(\m a)$ and $F\equiv\phi/\m$ have been introduced. Let us note that $A$ and $F$ vary in the interval $]-\infty,+\infty[$ and that, at fixed $A$ and for standard potentials, the inflaton wave function is normalizable in the usual sense,
\[
\int_{-\infty}^{+\infty}{\rm d}F\,|\Psi|^2<+\infty.
\]
Despite its compact form, Eq.~(\ref{wdwG}) admits exact solutions only for a limited class of potentials. The same situation occurs at the classical level.

\subsection{Constant potential case}
If we consider a constant inflaton potential $V=\m^4\lambda$, the WdW equation can be easily solved by the separation of variables. An inflaton field evolving under a constant potential cannot reproduce the observed CMB spectrum and therefore does not represent, by itself, a realistic model of inflation. Nonetheless, this case is frequently considered in the literature as a toy model, since it captures several features of the inflationary evolution while admitting exact solutions for both the background dynamics and the associated MS equation. Moreover, in recent years, inflaton potentials featuring a plateau or an inflection point have been extensively investigated as possible mechanisms for the production of primordial black holes as dark matter~\cite{SasakiEtAl2018,BallesterosTaoso2018}. For this reason, the constant potential case may still provide some useful physical insight.
We briefly review the classical dynamics of this system, starting from the Klein--Gordon equation
\be{kgeq}
\ddot{\phi}+3H\dot{\phi}=0\,,
\ee
where the dot denotes differentiation with respect to cosmic time. This equation can be readily integrated to give
\[
\dot{\phi}=\frac{\m^2P}{a^3}\,,
\]
where $P$ is an integration constant satisfying $P=\pi_\phi/\m^2$. Correspondingly, the Hamiltonian constraint~(\ref{hamG}), which coincides with the Friedmann equation, becomes
\be{hamG3}
-a^2\pi_a^2+P^2+2a^6\m^6\lambda=0\,.
\ee
Different values of $P$ correspond to different background evolutions. The case $P=0$ describes a pure de Sitter spacetime with $H^2_{\rm dS}={\rm const}$. Solutions with $P\neq0$ asymptotically approach the de Sitter attractor, with 
\be{dsappMC}
H^2-H^2_{\rm dS}\sim P^2/a^6. 
\ee
For sufficiently large values of $|P|$, the phase of accelerated expansion is preceded by a kination-dominated era.\\
In the quantum framework, upon factorizing the inflaton and gravitational wave functions as follows
\be{sepcon}
\Psi(A,F)=\re^{iPF}\psi_P(A)
\ee
one obtains
\be{wdwGconst}
\left(\partial_A^2+P^2+2\re^{6A}\lambda\right)\psi_P(A)=0\,,
\ee
which has the general solution
\be{gsolcon}
\psi_P(A)=c_1H_\nu^{(1)}(z)+c_2H_\nu^{(2)}(z)
\ee
where $\nu=iP/3$, $z=\sqrt{2\lambda/9}\re^{3A}$, $c_1$ and $c_2$ are arbitrary integration constants, and $H_\nu^{(1,2)}$ are Hankel functions.

The general solution of the WdW equation is an arbitrary superposition of the form
\be{gensolcon}
\Psi(A,F)=\int_{-\infty}^{+\infty}\rd P\,\re^{iPF}
\left[c_1f_1(P)H_\nu^{(1)}(z)+c_2f_2(P)H_\nu^{(2)}(z)\right].
\ee

In what follows we make use of the following integral representation of the Hankel function
\be{intrepr}
H_\nu^{(1)}(z)=\frac{{\rm e}^{-i\pi\nu/2}}{i\pi}
\int_{-\infty}^{+\infty}\rd t\,
\re^{iz\cosh t-\nu t}.
\ee

Starting from (\ref{intrepr}) and in the large-$z$ limit, which also corresponds to the large-$A$ limit, one can apply the saddle-point approximation and easily derive the standard asymptotic expansion. Indeed, the dominant contribution to the integral arises from the neighborhood of the minimum of $\cosh t$, namely $t=0$. Expanding $\cosh t\simeq1+t^2/2$ and properly deforming the integration contour around $t=0$, one finds
\ba{intapprox}
\!\!\!\!\!\!\!\!\!\!H_\nu^{(1)}(z)&\!\!\!\!\sim\!\!\!\!&
\frac{\re^{-i\pi\nu/2}}{i\pi}
\int_{-\infty}^{+\infty}\rd t\,
\re^{iz+izt^2/2-\nu t}\nonumber\\
\!\!\!\!&\!\!\!\!=\!\!\!\!&
\frac{\re^{iz-i\pi\nu/2-i\pi/4}}{\pi}
\int_{-\infty}^{+\infty}\rd\tau\,
\re^{-z\tau^2/2-\re^{i\pi/4}\nu\tau}
=\sqrt{\frac{2}{\pi z}}
\re^{iz-i\pi\nu/2-i\pi/4+i\frac{\nu^2}{2z}}
\ea
which is the result of the quadratic saddle-point approximation. Compared to the usual large-$z$ expansion of the Hankel function, the contribution $-i/(8z)$ is missing because it can only be obtained by retaining higher-order terms beyond the quadratic expansion used in the saddle-point approximation. We have retained the linear term in $\nu t$ in order to keep the nontrivial dependence on $P$ explicit.

In the same limit, and to leading order, one has
\be{momAcon}
-i\partial_aH_\nu^{(1)}(z)=
\frac{\rd z}{\rd a}
\frac{{\rm e}^{-i\pi\nu/2}}{i\pi}
\int_{-\infty}^{+\infty}\rd t\,
\cosh t\,\re^{iz\cosh t-\nu t}
\stackrel{a\m\gg1}{\longrightarrow}
\sqrt{2\lambda}\,\m^3a^2H_\nu^{(1)}(z).
\ee
where $\sqrt{2\lambda}\,\m^3a^2=\pi_a^{\rm cl}$ corresponds to the classical Friedmann equation,
$H=-\sqrt{2\lambda\,\m^2}$, describing a contracting universe. Conversely, the quantum evolution of an expanding universe is associated with the second Hankel function, $H_\nu^{(2)}(z)$.
\subsection{Gaussian wave-packet solution}
\begin{figure}[t]
  \centering
  \includegraphics[width=0.45\textwidth]{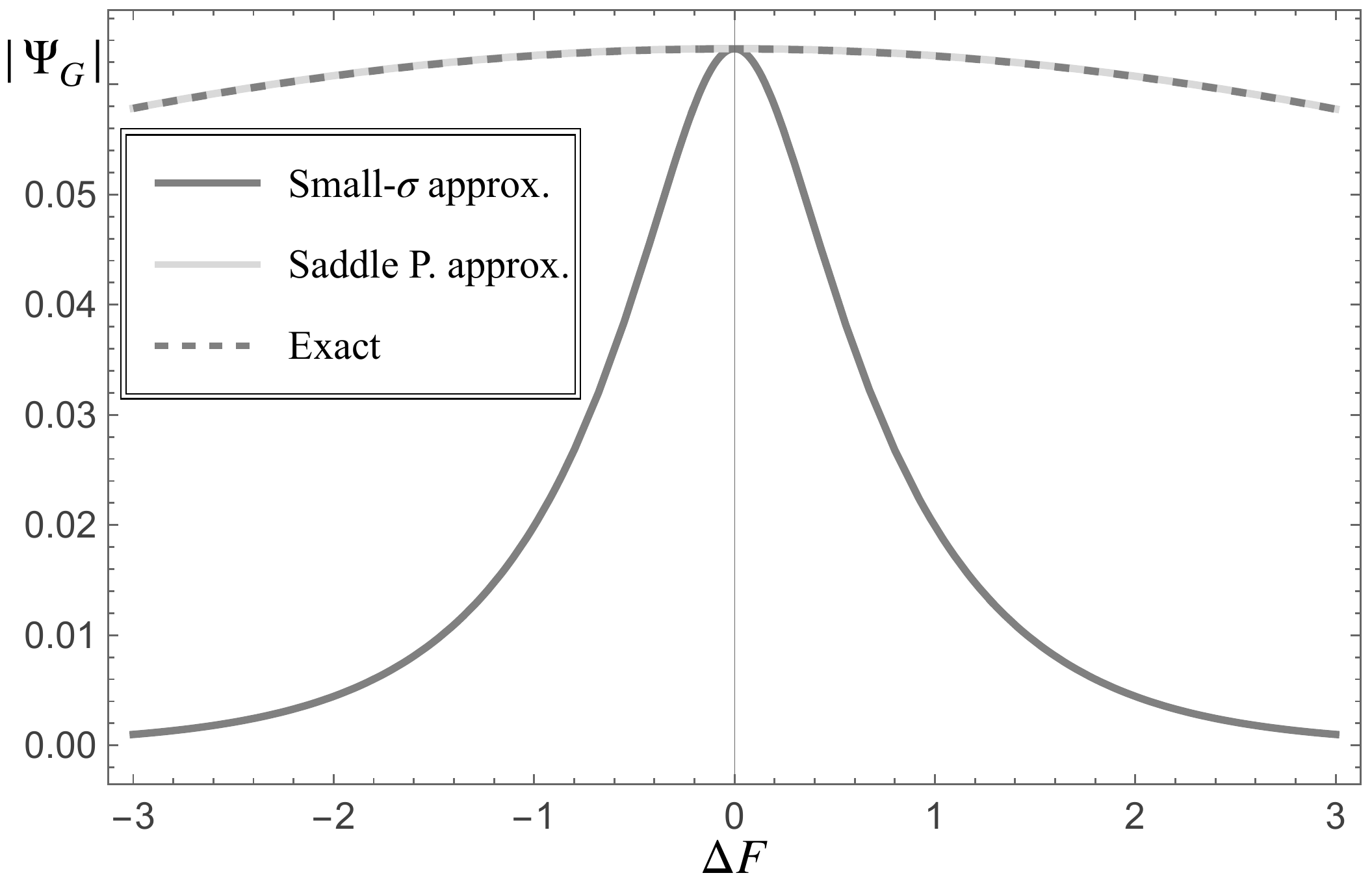}
  \includegraphics[width=0.453\textwidth]{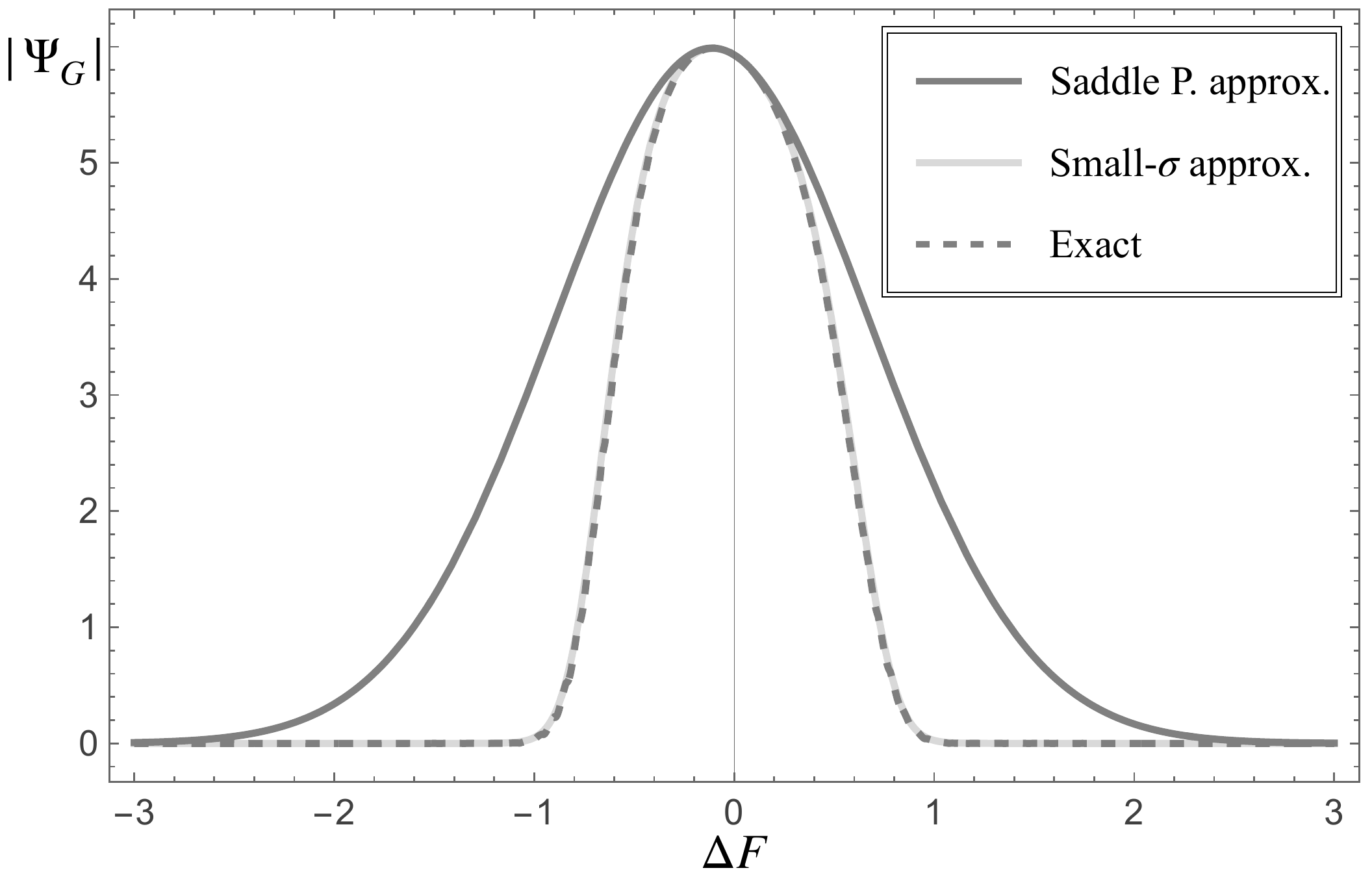}
  \caption{Left: comparison between the exact (numerical) Gaussian wave-packet modulus $|\Psi_G|$ (dashed curve) and the analytic approximations (\ref{modzlarge}) (light gray) and (\ref{modsigmasmall}) (gray) for $z=20$, $P_0=2$, and $\sigma=10$. Right: analogous comparison for $z=2$, $P_0=2$, and $\sigma=1/20$, where the colors of the two approximations are interchanged in order to better visualize the dashed curve, which is almost completely superimposed on one of the analytic curves. The vertical line marks $F=F_0$.\label{fig1}}
\end{figure}

Let us now consider the following Gaussian-weighted superposition of expanding Hankel functions:
\ba{gaus1}
\Psi_G(A,F)&=&N\int_{-\infty}^{+\infty}\rd P\,\re^{i P (F-F_0)-\sigma^2(P-P_0)^2} \re^{\pi P/6}H_\nu^{(2)}(z)\nonumber\\
&=&-\frac{N}{i\pi}\iint_{-\infty}^{+\infty}\rd P\,\rd t\,\re^{i P (F-F_0)-\sigma^2(P-P_0)^2}  \,\re^{-i z \cosh t+i P t/3}\nonumber\\
&=&-\frac{N}{i\sqrt{\pi \sigma^2}}\int_{-\infty}^{+\infty}\,\rd t\,\re^{-\frac{\paq{t+3(F-F_0)}^2}{36\sigma^2}+\frac{iP_0}{3}\paq{t+3(F-F_0)}} \,\re^{-i z \cosh t}.
\ea
Here, $N$ is a normalization factor, which will be included explicitly in the Gaussian wave packet from now on. It is fixed, at each value of $A$, by the condition
\be{normN}
\int_{-\infty}^{+\infty}\rd F\,|\Psi_G(A,F)|^2=1\,,
\ee
and is therefore generally a function of $A$. This normalization is required when calculating expectation values. However, because an $A$-dependent prefactor does not in general preserve the Wheeler-DeWitt equation, it is the unnormalized wave packet $\Psi_G(A,F)/N(A)$ that constitutes the exact solution of the WdW equation. \\
The last integral can be evaluated in two different regimes (see Fig. \ref{fig1} for a comparison between the analytical approximations employed and the exact expressions).
When $z\gg 1$ and $\sigma$ is sufficiently large, the integral can be evaluated by the saddle-point approximation, following the same procedure used in (\ref{intapprox}). In this case one has
\be{zlarge}
\Psi_G=N \frac{6 \,i\,\re^{-i\pi/4}}{\sqrt{-i+18 z \sigma^2}}\re^{\frac{-2i\sigma^2\paq{P_0^2+18 (F-F_0) P_0 z-18z^2}+z\paq{2+9(F-F_0)^2}}{2i-36z\sigma^2}}
\ee
and
\be{modzlarge}
|\Psi_G|=\frac{6|N|\re^{-\frac{\paq{P_0+9(F-F_0)z}^2\sigma^2}{1+324z^2\sigma^4}}}{\pa{1+324z^2\sigma^4}^{1/4}}.
\ee
The resulting expression is a Gaussian peaked at
\be{traj1}
F=F_0-\frac{P_0}{3\sqrt{2\lambda}} \re^{-3A},
\ee
i.e. along the classical solution. 

In contrast, in the small-$\sigma$ limit, the dominant contribution to the integral comes from the neighborhood of $t=-3(F-F_0)$. Expanding the exponent to second order around $t=-3(F-F_0)$ and performing the integral, one obtains
\ba{sigmasmall}
\Psi_G\!\!\!\!&\simeq&\!\!\!\! \frac{6 N\re^{i\pi/4}}{\sqrt{-i+18 z \sigma^2 \cosh3(F-F_0)}}\nonumber\\
&&\!\!\!\!\times\re^{\frac{2i\sigma^2\paq{P_0^2+6 P_0 z\sinh 3(F-F_0)-27/2z^2-9/2z^2\cosh6(F-F_0)}-2z\cosh3(F-F_0)}{-2i+36z\sigma^2\cosh3(F-F_0)}}
\ea
and
\be{modsigmasmall}
|\Psi_G|=\frac{6 |N|\re^{-\frac{\paq{P_0+3z\sinh3(F-F_0)}^2\sigma^2}{1+324z^2\sigma^4\cosh^23(F-F_0)}}}{\paq{1+324z^2\sigma^4\cosh^23(F-F_0)}^{1/4}}.
\ee

The shape of the packet now is no longer purely Gaussian. Its peak can be estimated by calculating the minimum of the argument of the exponent, yielding
\be{traj2}
F=F_0-\frac{1}{3}\operatorname{arcsinh} \pa{\frac{P_0}{3z}}.
\ee
This trajectory deviates from the classical one until $F$ approaches its attractor value $F_0$. Thus, the wave packet does not follow the classical trajectory until $z$ becomes sufficiently large. At this point the present approximation is no longer valid, and one recovers the previous result (\ref{modzlarge}).

One can finally calculate the normalization factor $N$ from the exact integral
\be{norm2}
N^{-2}=\sqrt{\frac{2}{\pi\sigma^2}}\iint_{-\infty}^{+\infty}\rd t\rd \bar t \re^{-\frac{\pa{t-\bar t}^2}{72\sigma^2}+i \frac{P_0}{3}(t-\bar t)-iz\pa{\cosh t-\cosh\bar t}}
\ee
which can be evaluated using either the Gaussian or the saddle-point approximation. In particular, in the saddle-point approximation one has $N^{-2}=\sqrt{8\pi}/(z\sigma)$.

\section{Induced Gravity}

A procedure analogous to that illustrated in the previous section can be followed in the induced-gravity (IG) framework, where Newton's constant is dynamically generated by the expectation value of the inflaton field [namely by replacing $\m^2/6\rightarrow U=\xi\phi^2$ in the action (\ref{genact})]. General solutions of the WdW equation for IG 
\be{wdwIG}
\paq{\frac{1}{12\xi }\partial_A^2+\partial_A\partial_F-\frac{1}{2}\partial_F^2+a^6\pa{1+6\,\xi}\phi^2V}\Psi(a,\phi)=0,
\ee
with $F\equiv\ln(\phi/\m)$, have been presented in \cite{KamenshchikTronconiVenturi2019IG} for $V=\lambda \m^{4-n}\phi^n$. In this context, one finds a set of exact factorized solutions of (\ref{wdwIG}) given by
\be{sol1IG}
\Psi_\mu(a,\phi)=\left(\frac{\phi}{\m}\right)^\mu\bar\chi_\mu(x)
\ee
with $x=a^3\phi^{\frac{n+2}{2}}$. The first factor in (\ref{sol1IG}) can be rewritten as a plane wave in the variable $F$, with $\mu=iP$ purely imaginary:
\be{sol1pwIG}
\Psi_\mu(a,\phi)=\re^{iPF}\bar\chi_\mu(x).
\ee
The second factor, $\bar\chi_\mu(x)$, satisfies the following second-order equation:
\be{wdwans2}
\paq{1-\frac{\xi^2\pa{n-4}^2}{\Gamma^2}}\frac{\rd^2\bar \chi}{\rd \pa{\ln x}^2}-4 i P\frac{\xi^2\pa{n-4}}{\Gamma^2}\frac{\rd\bar \chi}{\rd \ln x}+W_2(x)\bar\chi=0
\ee
where
\be{Wdef}
W_2(x)\equiv\paq{\frac{4\xi^2P^2}{\Gamma^2}+\frac{4}{3}\xi\lambda \m^{4-n}x^2},
\ee
and $\Gamma=\sqrt{6\xi(1+6\xi)}$.
\subsection{Quartic potential}
Let us consider the simplest case, namely the quartic potential with $n=4$. For this potential the background cosmological evolution resembles that of a minimally coupled inflaton in a constant potential. In particular, the system admits a de Sitter attractor, and scalar-field trajectories close to this attractor satisfy $\dot \phi\sim P/a^3$ where $P$ is an integration constant. In contrast to the minimally coupled case, however, the deviation from the de Sitter expansion behaves as 
\be{dsattNMC}
H^2-H^2_{\rm dS}\sim P/a^3.
\ee
At the quantum level, in this case, one can define $x=(a\phi)^3=\re^{3(A+F)}\equiv \re^{3y}$ and the ansatz (\ref{sol1pwIG}) leads to the following equation for $\bar\chi_\mu$
\be{wdwans4}
\frac{\rd^2\bar \chi_\mu}{\rd y^2}+\paq{\frac{36\xi^2P^2}{\Gamma^2}+ 12\xi\,\lambda\,\re^{6y}}\bar\chi_\mu=0
\ee
which is formally similar to (\ref{wdwGconst}) and has the general solution
\be{gsolig}
\bar \chi_\mu(x)=c_1 H_\nu^{(1)}(z)+c_2 H_\nu^{(2)}(z)
\ee
where $\nu=2i\xi P/\Gamma$, $z=\sqrt{4\xi\lambda/3}\,\re^{3y}$, and $c_1$ and $c_2$ are arbitrary integration constants. A possible Gaussian solution is then
\be{gaussIG}
\Psi_G=N\int_{-\infty}^{+\infty}{\rd P} \re^{i P (F-F_0)-\sigma^2\pa{P-P_0}^2}\re^{-i\pi \nu/2}H_{\nu}^{(2)}(z)
\ee
and, if one redefines $P=\Gamma/(6\xi)\,\widetilde P$ and $F=6\xi/\Gamma\, \widetilde F$, the expression (\ref{gaussIG}) takes the form
\be{gaussIG2}
\Psi_G=N\frac{\Gamma}{6\xi}\int_{-\infty}^{+\infty}{\rd \widetilde P}\re^{i\widetilde P\pa{{\widetilde F}-{\widetilde F}_0}-\sigma^2\Gamma^2/(6\xi)^2\pa{\widetilde{P}-\widetilde{P}_0}^2}\re^{\pi \widetilde P/6}H_{i\widetilde P/3}^{(2)}(z)
\ee
which is formally identical to (\ref{gaus1}), provided one defines $\widetilde \sigma=\sigma \Gamma/(6\xi)$ and $\widetilde N=N \Gamma/(6\xi)$. The approximations obtained previously from (\ref{gaus1}) can then be straightforwardly generalized to the present case through these simple substitutions.

Using the integral representation adopted in the previous section one finds
\ba{psiGig}
\Psi_G(A,F)&=&-\frac{N}{i\pi}\frac{\Gamma}{6\xi}\iint_{-\infty}^{+\infty}\rd {\widetilde P}\,\rd t\,\re^{i {\widetilde P} ({\widetilde F}-{\widetilde F}_0)-{\widetilde\sigma}^2({\widetilde P}-{\widetilde P}_0)^2}  \,\re^{-i z \cosh t+i {\widetilde P} t/3}\nonumber\\
&=&-\frac{\widetilde N}{i\sqrt{\pi {\widetilde\sigma}^2}}\int_{-\infty}^{+\infty}\,\rd t\,\re^{-\frac{\paq{t+3(\widetilde F-\widetilde F_0)}^2}{36{\widetilde\sigma}^2}+\frac{i{\widetilde P}_0}{3}\paq{t+3(\widetilde F-\widetilde F_0)}} \,\re^{-i z \cosh t}.
\ea
Here $z=\sqrt{4\xi\lambda/3}\,\re^{3(A+F)}$ and $F$ varies over the real axis. Therefore, the saddle-point approximation is not uniformly valid throughout the integration domain in $F$. However, the Gaussian approximation can be accurate for small $\widetilde\sigma$. Varying $F$, one finds that the maximum of the integral occurs when
\be{max}
F-F_0=-\frac{2\xi}{\Gamma}\operatorname{arcsinh} \pa{\frac{2\xi P_0}{\Gamma z}}. 
\ee


\section{Inflationary perturbations}

Let us now illustrate how perturbations can be taken into account in the case of a minimally coupled inflaton.\footnote{For the IG case and more general nonminimally coupled inflaton fields, the general treatment of perturbations is very similar and is presented in \cite{KamenshchikTronconiVenturi2020Nonminimal,SteinwachsVanDerWild2018,SteinwachsVanDerWild2019}.}

When perturbations are included, the matter-gravity Hamiltonian (\ref{hamG}) must be modified accordingly,
$\mathcal{H}\rightarrow \mathcal{H}+\sum_kH_k\equiv H_{\rm tot}$,
where
\be{Hkdef}
H_k\equiv \frac{1}{2}\pa{\pi_k^2+\omega_k^2v_k^2},
\ee
$v_k$ is the Mukhanov-Sasaki variable, $\pi_k\equiv v'_k$, and the frequency squared $\omega_k^2\equiv k^2-z''_{\rm MS}/z_{\rm MS}$ classically depends on the evolution of the homogeneous degrees of freedom $(a,\phi)$ through $z_{\rm MS}$. Let us note that, for simplicity, and without loss of generality, we consider a single perturbation mode. The same procedure and identical results are obtained when the full tower of $k$ modes is taken into account (see Ref.~\cite{KTV2015} for details).

Upon canonically quantizing both the homogeneous degrees of freedom and the perturbation variable $v_k$, the Hamiltonian constraint becomes the WdW equation
\be{wdwp}
\hat H_{\rm tot}|\bar{\Psi}(a,\phi,\{v_k\})\rangle=0.
\ee
Here, the variables $(a,\phi)$, which enter $\omega_k$, are treated classically. This approximation amounts to neglecting the contribution of the quantum fluctuations of the homogeneous quantities to the evolution of perturbations. Indeed, one is interested in calculating the power spectrum of $v_k$, which, to leading order, is $\langle 0|\hat v_k^2|0\rangle\propto \hbar$ in the standard semiclassical formalism. The inclusion of the quantum fluctuations of the homogeneous quantities, for example $\av F=F_{\rm cl}+\mathcal{O}(\hbar)$, would give an $\mathcal{O}\pa{\hbar^2}$ contribution to the spectrum, which will be neglected in the following.

The corresponding WdW equation can therefore be written in the compact form
\be{wdwGp}
\paq{\eta^{IJ} \partial_I\partial_J+2\m^{-4}\re^{6A}V+\re^{2A}\pa{\hat \pi_k^2+\omega_k^2\hat v_k^2}} \bar\Psi(A,F,v_k)=0
\ee
where $I,J=1,2$ correspond, respectively, to $A,F$, and $\eta^{IJ}={\rm diag}\pag{1,-1}$.
We consider factorized solutions of the form
\be{barpsidec}
|\bar\Psi(A,F,v_k)\rangle=\widetilde\Psi(A,F)|\widetilde\chi_k(A,F,v_k)\rangle
\ee
where $|\widetilde\chi_k(A,F,v_k)\rangle$ is the properly normalized wave function for $v_k$.\footnote{If the full spectrum of perturbations is retained, the replacement $|\widetilde\chi_k\rangle\rightarrow \prod_k|\widetilde\chi_k\rangle$ must be performed.}
Equation (\ref{wdwGp}) becomes
\ba{wdwGpExpanded}
&&\eta^{IJ}\paq{\pa{ \partial_I\partial_J\widetilde\Psi}|\widetilde\chi_k\rangle+2\pa{\partial_I\widetilde\Psi}\partial_J|\widetilde\chi_k\rangle+\widetilde\Psi\partial_I\partial_J|\widetilde\chi_k\rangle}\nonumber\\
&&+2\m^{-4}\re^{6A}V\,\widetilde\Psi|\widetilde\chi_k\rangle+\re^{2A}\widetilde\Psi\pa{\hat \pi_k^2+\omega_k^2\hat v_k^2} |\widetilde\chi_k\rangle=0
\ea
and, upon contracting this equation with $\langle \widetilde\chi_k|$, one obtains the equation for the wave function of the homogeneous degrees of freedom:
\be{wdwGpH}
\pag{\eta^{IJ}\paq{\partial_I\partial_J+2 \av{\widetilde{\partial_J}}_k\partial_I+ \av{\widetilde{\partial_I\partial_J}}_k}
+2\m^{-4}\re^{6A}V+2\re^{2A}\av {\widetilde{\hat H_k}}_k}\widetilde\Psi=0,
\ee
where $\av{\widetilde{\hat O}}_k\equiv \langle\widetilde\chi_k|\hat O|\widetilde\chi_k\rangle$.

On multiplying (\ref{wdwGpH}) by $|{\widetilde \chi}_k\rangle$ and subtracting the resulting equation from (\ref{wdwGpExpanded}), one obtains the equation for the wave function of $v_k$:
\ba{wdwGpvk}
&&\eta^{IJ}\paq{2\pa{\partial_I\widetilde\Psi}\widetilde\Psi^{-1}\pa{\partial_J-\av{\widetilde{\partial_J}}_k}|\widetilde\chi_k\rangle+\pa{\partial_I\partial_J-\av{\widetilde{\partial_I\partial_J}}_k}|\widetilde\chi_k\rangle}\nonumber\\
&&+2\re^{2A}\pa{\hat H_k-\av{\widetilde{\hat H_k}}_k} |\widetilde\chi_k\rangle=0.
\ea
The above derivation corresponds to the standard Born-Oppenheimer decomposition of the WdW equation, in which the homogeneous degrees of freedom are treated as slow variables, whereas the perturbations are treated as fast variables~\cite{BornOppenheimer1927,BertoniFinelliVenturi1996,KTV2018BO}.
One can now redefine $\widetilde\Psi$ and $|\widetilde\chi_k\rangle$ having opposite phases, i.e.
\be{rephase}
\widetilde \Psi=\re^{i\theta}\Psi,\quad |\widetilde\chi_k\rangle=\re^{-i\theta}|\chi_k\rangle
\ee
with $\theta=\theta(A,F)$, thus obtaining
\be{dIchik}
\partial_I |\widetilde\chi_k\rangle=\re^{-i\theta}\paq{\pa{\partial_I |\chi_k\rangle}-i\pa{\partial_I\theta} |\chi_k\rangle},
\ee
\be{ddIchik}
\partial^2_I |\widetilde\chi_k\rangle=\re^{-i\theta}\pag{\pa{\partial_I^2 |\chi_k\rangle}-2i\pa{\partial_I\theta} \pa{\partial_I|\chi_k\rangle}-\paq{i\pa{\partial_I^2\theta}+\pa{\partial_I\theta}^2}|\chi_k\rangle}
\ee
and
\be{dIpsi}
\partial_I \widetilde\Psi=\re^{i\theta}\paq{\pa{\partial_I \Psi}+i\pa{\partial_I\theta} \Psi},
\ee
\be{ddIpsi}
\partial^2_I \widetilde\Psi=\re^{i\theta}\pag{\pa{\partial_I^2 \Psi}+2i\pa{\partial_I\theta} \pa{\partial_I\Psi}+\paq{i\pa{\partial_I^2\theta}-\pa{\partial_I\theta}^2}\Psi}.
\ee

On contracting (\ref{dIchik}) and (\ref{ddIchik}) with $\langle\widetilde\chi_k|=\re^{i\theta}\langle\chi_k|$, one further obtains
\be{dIav}
\av{\widetilde{\partial_I}}_k =\av{\partial_I}_k-i\partial_I\theta,
\ee
\be{ddIav}
\av{\widetilde{\partial^2_I}}_k=\av{\partial_I^2}_k-2i\pa{\partial_I\theta} \av{\partial_I}_k-\paq{i\pa{\partial_I^2\theta}+\pa{\partial_I\theta}^2}.
\ee

If one can choose $\theta$ in such a way that $\av{\partial_I}_k=0$ and
\be{sim2}
\langle\partial_I\chi_k|\partial_I \chi_k\rangle+\langle \chi_k|\partial_I^2\chi_k\rangle=0,
\ee
then the equations for the homogeneous inflaton-gravity system (\ref{wdwGpH}) and for the perturbations (\ref{wdwGpvk}) simplify to
\be{wdwGpHsim}
\eta^{IJ} \partial_I\partial_J\Psi
+2\m^{-4}\re^{6A}V\,\Psi=-2\re^{2A}\Psi\av {\hat H_k}_k-\eta^{IJ}\Psi\av{{\partial_I\partial_J}}_k,
\ee
\be{wdwGpvksim}
\eta^{IJ}\frac{\partial_I\Psi}{\Psi}\partial_J|\chi_k\rangle+\re^{2A}\pa{\hat H_k-\av{\hat H_k}_k} |\chi_k\rangle=\frac{\eta^{IJ}}{2}\pa{\av{\partial_I\partial_J}_k-\partial_I\partial_J}|\chi_k\rangle.
\ee
The right-hand sides (rhs's) of these equations can be neglected. Indeed, the rhs of Eq.~(\ref{wdwGpHsim}) describes the backreaction of the perturbations on the homogeneous inflaton-gravity system, which is also usually neglected in the conventional semiclassical approach. Likewise, the rhs of Eq.~(\ref{wdwGpvksim}) contains the nonadiabatic corrections to the evolution of $v_k$. These corrections have been studied in previous works \cite{AlberghiCasadioTronconi2006,KTV2013,KTV2014,KTV2015,KTV2016,BrizuelaKieferKramer2016dS,BrizuelaKieferKramer2016SR,KamenshchikTronconiVenturi2020Nonminimal}, are always present within the Born-Oppenheimer treatment, and are typically tiny, being of order $\sim (\hbar/\m)^2$.
The final resulting approximate system of equations is therefore
\be{syseqs}
\left\{
\begin{array}{l}
\pa{\eta^{IJ} \partial_I\partial_J+2\m^{-4}\re^{6A}V}\,\Psi=0\\
\eta^{IJ}\frac{\partial_I\Psi}{\Psi}\partial_J|\chi_k\rangle+\re^{2A}\pa{\hat H_k-\av{\hat H_k}_k} |\chi_k\rangle=0
\end{array}
\right.
\ee
where the first equation coincides with the homogeneous WdW equation (\ref{wdwG}). The second equation can be recast as a Schwinger-Tomonaga equation for the wave function $|\chi_k\rangle$, which is equivalent to the Mukhanov-Sasaki equation. This becomes apparent once a suitable notion of time evolution is introduced.
\subsection{Emergence of time}
We now show how time emerges in the equation for the perturbations~\cite{BertoniFinelliVenturi1996,KieferSingh1991,KTVV2019Time,KTV2021BO,ChataignierKramer2021}. For simplicity we examine the constant potential case which has been discussed in detail in Sec.~II A. Starting from the second approximate equation in (\ref{syseqs})
\be{timems}
\eta^{IJ}\frac{\partial_I\Psi}{\Psi}\partial_J|\chi_k\rangle+\re^{2A}\pa{\hat H_k-\av{\hat H_k}_k} |\chi_k\rangle=0
\ee
and the solution to the first,
\be{defPsi}
\Psi=N \re^{i P F} H_\nu^{(2)}(z),
\ee
at leading order in the large-$z$ expansion one has 
\be{cltime}
\eta^{IJ}\frac{\partial_I\Psi}{ \Psi}\partial_J|\chi_k\rangle=-i\pa{\sqrt{2\lambda}\re^{3A}\partial_A+P\partial_F}|\chi_k\rangle
\ee
where $\sqrt{2\lambda}\re^{A}=A_{\rm cl}'$ and $\re^{-2A}P=F'_{\rm cl}$. Thus 
\be{cltime2}
\eta^{IJ}\frac{\partial_I \Psi}{ \Psi}\partial_J|\chi_k\rangle=-i\re^{2A}\paq{A'_{\rm cl}(\eta)\partial_A+F'_{\rm cl}(\eta)\partial_F}|\chi_k\rangle=-i\re^{2A}\frac{\rd}{\rd \eta}|\chi_k\rangle
\ee
is the total (conformal) time derivative along the classical trajectory and Eq. (\ref{timems}) can be rewritten as 
\be{ScTom}
i\frac{\rd}{\rd \eta}|\chi_k\rangle_s=\hat H_k(\eta)|\chi_k\rangle_s
\ee
where
\be{chis}
|\chi_k\rangle_s=|\chi_k\rangle\re^{-i\int^{\eta}\rd \eta'\av{\hat H_k(\eta')}_k}
\ee
which is the Schr\"odinger equation (or Schwinger-Tomonaga equation) corresponding to the MS equation. Once time evolution (in this case parametrized by the conformal time) is introduced, an arbitrary reparametrization of time may then be used to solve Eq. (\ref{ScTom}). In particular, $A$, which physically corresponds to the classical number of e-folds, up to an additive constant, can be used to parametrize the evolution of $|\chi_k\rangle$.
Let us now note that the cancellations leading to Eq.~(\ref{syseqs}) are a consequence of a suitable phase redefinition yielding $\av{\partial_I}_k=0$, or, equivalently,
\be{dIav2}
\partial_A\theta=i\av{\widetilde{\partial_A}}_k,\;{\rm and}\;\partial_F\theta=i\av{\widetilde{\partial_F}}_k.
\ee
Such a phase $\theta$ exists only if
\be{exdiff}
\partial_F\av{\widetilde{\partial_A}}_k=\partial_A\av{\widetilde{\partial_F}}_k\Leftrightarrow {\rm Im}\langle\partial_F \widetilde\chi_k|\partial_A\widetilde \chi_k\rangle=0.
\ee
Since the quantum fluctuations entering $\omega_k$ are neglected, the latter is evaluated along the classical trajectory. One therefore has
$F=F_{\rm cl}(A)$ and $\partial_F=\pa{\rd F_{\rm cl}/\rd A}^{-1}\partial_A$. The condition (\ref{exdiff}) is therefore automatically satisfied, and
\be{exdiff2}
\langle\partial_F \widetilde\chi_k|\partial_A\widetilde \chi_k\rangle=\pa{\rd F_{\rm cl}/\rd A}^{-1} \left\||\partial_A\widetilde \chi_k\rangle\right\|^2.
\ee

The resulting solution of the WdW equation (\ref{wdwp}) is
\be{solwdwtot}
\bar\Psi=\Psi(A,F)|\chi(A,v_k)\rangle=\Psi(A,F)|\chi(A,v_k)\rangle_s \re^{i\int^{\eta}\rd \eta'\av{\hat H_k(\eta')}_k}
\ee
where backreaction and nonadiabatic effects have been neglected. Owing to the linearity of the WdW equation, any superposition of the approximate solutions obtained through the above procedure is a solution to the same order of approximation. Thus
\be{solwdwtoti}
\bar\Psi=\sum_i c_i\Psi_i(A,F)|\chi_i(A,v_k)\rangle
\ee
and, in particular, for the Gaussian wave-packet solution (\ref{gaus1}), $i\rightarrow P$ and
\be{solwdwtotp}
|\bar\Psi(A,F,v_k)\rangle=N\int_{-\infty}^{+\infty}\rd P \re^{iP(F-F_0)-\sigma^2(P-P_0)^2}\re^{\pi P/6}H_\nu^{(2)}(z)|\chi_P(A,v_k)\rangle.
\ee
\section{Power Spectrum}
Let us now calculate the spectrum of the MS variable, $\av{v_k^2}$. Within the standard semiclassical approach there is a single classical solution describing the homogeneous background, and
\be{defPSsc}
P_{v}\pa{k;P_0}\equiv \av{v_k^2}_A= \langle \chi_{P_0}(A,v_k)|\hat v_k^2|\chi_{P_0}(A,v_k)\rangle\equiv \int_{-\infty}^{+\infty}\!\!\!\!{\rd v_k} \left|\chi_{P_0}\right|^2 v_k^2
\ee
where $P_0$ labels the particular classical trajectory and $A$ plays the role of the time variable. In the fully quantized description, the homogeneous inflaton field is instead described by a wave function with a Gaussian distribution peaked around $P_0$. Let us note that $P_{v}(k)$ is related to the power spectrum of curvature perturbations $\mathcal R_k$ by the relation~\cite{MukhanovFeldmanBrandenberger1992}
\be{PSdef}
\mathcal{P}_{\mathcal R}=\frac{k^3}{2\pi^2}\left.\frac{P_v(k)}{z_{\rm MS}^2}\right|_{A_{\rm end}}.
\ee
where $A_{\rm end}$ is the value of $A$ at the end of inflation and $z_{\rm MS}$ is the time-dependent function entering $\omega_k^2$. 
If $A$ is used as the classical time variable, one has
\be{defPS}
\av{v_k^2}_A= \langle \bar\Psi(A,F,v_k)|\hat v_k^2|\bar\Psi(A,F,v_k)\rangle\equiv \iint_{-\infty}^{+\infty}\!\!\!\! \rd v_k \rd F \left|\bar\Psi\right|^2 v_k^2
\ee
and the inflationary power spectrum is now obtained after integrating also over the homogeneous degree of freedom $F$. More explicitly, for the solution (\ref{solwdwtotp}) one finds
\ba{PS0}
\av{v_k^2}_A\!\!\!\!&=\!\!\!\!&N^2\iiint _{-\infty}^{+\infty}\!\!\!\!\rd F \rd P\rd \bar P  \re^{i(P-\bar P)(F-F_0)-\sigma^2(\bar P-P_0)^2-\sigma^2(P-P_0)^2}\re^{\pi \pa{\bar P+P}/6}\nonumber\\
&&\times\pa{H_{i\bar P/3}^{(2)}(z)}^*H_{iP/3}^{(2)}(z)\langle\chi_{\bar P}(A,v_k)|\hat v_k^2|\chi_P(A,v_k)\rangle
\ea
where the integration over $F$ produces the Dirac $\delta$ distribution $2\pi \delta (P-\bar P)$. We observe that, once time has been introduced and the tiny quantum corrections are neglected, one recovers the semiclassical MS equation, in which the dependence on $F$ is replaced by its corresponding $P$-dependent classical trajectory $F_{\rm cl}(A;P)$. Performing also the integration over $\bar P$, one obtains
\ba{PS1}
\av{v_k^2}_A\!\!\!\!&=\!\!\!\!&2\pi N^2\int _{-\infty}^{+\infty}\!\!\!\!\rd P  \re^{-2\sigma^2(P-P_0)^2}\re^{\pi P/3}\nonumber\\
&&\times\pa{H_{\nu}^{(2)}(z)}^*H_{\nu}^{(2)}(z)\langle\chi_{P}(A,v_k)|\hat v_k^2|\chi_P(A,v_k)\rangle=\nonumber\\
&=\!\!\!\!&2\pi N^2\int _{-\infty}^{+\infty}\!\!\!\!\rd P  \re^{-2\sigma^2(P-P_0)^2}\re^{\pi P/3}\left|H_{\nu}^{(2)}(z)\right|^2P_{v}\pa{k;P}.
\ea

As expected, different background trajectories do not interfere because the MS equation is solved along each classical background trajectory independently. Nevertheless, describing the homogeneous inflaton by a Gaussian wave packet rather than by a classical c-number quantity has non-negligible consequences for the resulting power spectrum. The above expression can be further simplified in the large-$a$ limit (and the saddle-point approximation). Indeed, starting from the integral representation of the Hankel functions, Eq.~(\ref{intrepr}), one finds
\be{modH2}
\left|H_{\nu}^{(2)}(z)\right|^2=\iint_{-\infty}^{+\infty}\!\!\frac{\rd t\rd \bar t}{\pi^{2}}\re^{-\pi P/3}\re^{iP(t-\bar t)/3-iz\pa{\cosh t-\cosh \bar t}}\simeq\frac{2}{z\pi}\re^{-\pi P/3}
\ee
and Eq.~(\ref{PS1}) reduces to
\be{PS2GR}
\av{v_k^2}_A=\frac{4}{z}\pa{\sqrt{8\pi}/(z\sigma)}^{-1}\int _{-\infty}^{+\infty}\!\!\!\!\rd P  \re^{-2\sigma^2(P-P_0)^2}P_{v}\pa{k;P}
\ee
where the proper normalization factor has been used. For small $P\re^{-3A}$, the next-to-leading-order correction to $P_{v}$ is proportional to $P^2$ and one has
\be{PS2P}
\av{v_k^2}_A=\sqrt{\frac{2\sigma^2}{\pi}}\int _{-\infty}^{+\infty}\!\!\!\!\rd P  \re^{-2\sigma^2(P-P_0)^2}\pa{c_k^{(0)}\!\!+c_k^{(2)}P^2}=c_k^{(0)}\!\!+c_k^{(2)}\frac{1+4\sigma^2 P_0^2}{4\sigma^2}
\ee
As $\sigma$ decreases, the additional $P_0$-independent term $c_k^{(2)}/(4\sigma^2)$ grows, whereas the $P_0$-dependent term $c_k^{(2)}P_0^2$ is unchanged. We further note that the $k$ dependence of $c_k^{(0)}$ and $c_k^{(2)}$ may differ, leading to perturbative corrections in both the spectrum normalization and the spectral index.

If one repeats the same procedure in the IG case, the calculations are more cumbersome since $z=2\sqrt{\xi\lambda/3}\re^{3A+3F}$ and is a function of $F$. If we define, in order to maintain compact notation, 
\be{intvk}
f(A;P,\bar P)=\int_{-\infty}^{+\infty}\!\!\!\!{\rd v_k}v_k^2\;\chi_{\bar P}(A,v_k)^*\chi_P(A,v_k)\equiv \langle\chi_{\bar P}|\hat v_k^2|\chi_P\rangle
\ee
and integrate over $\phi$, which in IG varies in the interval $]0,+\infty[$, one has $\rd \phi=\m \re^{F}\rd F$.
\ba{PS2IG}
\av{v_k^2}_A\!\!\!\!&=\!\!\!\!&N^2\m \re^{F_0}\int _{-\infty}^{+\infty}\!\!\!\!\rd F \iint_{-\infty}^{+\infty} \!\!\!\!\rd t\rd \bar t \iint_{-\infty}^{+\infty} \!\!\!\!\rd P\rd \bar P  \re^{\paq{1+i(P-\bar P)}(F-F_0)}\nonumber\\
&&\times\re^{-\sigma^2(\bar P-P_0)^2-\sigma^2(P-P_0)^2-i z (\cosh t-\cosh \bar t)+2 i \xi Pt/\Gamma-2 i \xi \bar P\bar t/\Gamma}f(A;P,\bar P).
\ea
This expression cannot be simplified as we previously did for (\ref{PS0}). In the IG model with a quartic potential, the leading-order corrections to $P_v(k)$ are linear in $P\re^{-3A}$. Retaining only such corrections and using $\langle\chi_{\bar P}|\hat v_k^2|\chi_P\rangle^*=\langle\chi_{P}|\hat v_k^2|\chi_{\bar P}\rangle$, we have
\be{matrixP}
f(A;P,\bar P)=c_k^{(0)}\!+r_k^{(1)}(P+\bar P)\!+i s_k^{(1)}(P-\bar P)
\ee
and one can evaluate the integral (\ref{PS2IG}) in the small-$\sigma$ regime. Using the explicit form assumed for $f(A;P,\bar P)$ one can then proceed by integrating exactly over $P$ and $\bar P$ first. For small $\sigma$, one can then expand the hyperbolic cosine around the Gaussian peak to second order and proceed with the integration over $t$ and $\bar t$. One is then left with three integrals. The integral which multiplies $c_k^{(0)}$ is 
\be{normint}
I_0=\widetilde{\mathcal{N}}^2\int _{-\infty}^{+\infty}\!\!\!\!\rd F\re^{(F-F_0)}\frac{\exp\!\left[-\frac{2\xi^{2}\sigma^{2}\left(\frac{2P_{0}\xi}{\Gamma}+z\sinh\!\left(\frac{(F-F_{0})\Gamma}{2\xi}\right)\right)^{2}}{\Gamma^{2}\left(\frac{4\xi^{4}}{\Gamma^{4}}+z^{2}\sigma^{4}\cosh^{2}\!\left(\frac{(F-F_{0})\Gamma}{2\xi}\right)\right)}\right]}{\sqrt{\frac{4\xi^{4}}{\Gamma^{4}}+z^{2}\sigma^{4}\cosh^{2}\!\left(\frac{(F-F_{0})\Gamma}{2\xi}\right)}}
\ee
and defines the normalization factor $\widetilde{\mathcal{N}}$ ($I_0=1$). The integral multiplying $r_k^{(1)}$ is
\ba{in1r}
I_{1,r}&\!\!\!\!=&\!\!\!\!\widetilde{\mathcal{N}}^2\int _{-\infty}^{+\infty}\!\!\!\!\rd F\frac{\left[2P_{0}z^2\sigma^{4}\cosh^2\!\left(\frac{(F-F_{0})\Gamma}{2\xi}\right)-\frac{4\xi^{3}}{\Gamma^{3}}z\sinh\!\left(\frac{(F-F_{0})\Gamma}{2\xi}\right)\right]}{\left(\frac{4\xi^{4}}{\Gamma^{4}}+z^{2}\sigma^{4}\cosh^{2}\!\left(\frac{(F-F_{0})\Gamma}{2\xi}\right)\right)^{3/2}}\nonumber\\
&\!\!\!\!\times&\!\!\!\! \re^{(F-F_0)}\exp\!\left[-\frac{2\xi^{2}\sigma^{2}\left(\frac{2P_{0}\xi}{\Gamma}+z\sinh\!\left(\frac{(F-F_{0})\Gamma}{2\xi}\right)\right)^{2}}{\Gamma^{2}\left(\frac{4\xi^{4}}{\Gamma^{4}}+z^{2}\sigma^{4}\cosh^{2}\!\left(\frac{(F-F_{0})\Gamma}{2\xi}\right)\right)}\right]
\ea
and the one which multiplies $i s_k^{(1)}$ is
\ba{in1s}
I_{1,s}&\!\!\!\!=&\!\!\!\!i\,\widetilde{\mathcal{N}}^2\int _{-\infty}^{+\infty}\!\!\!\!\rd F\frac{\xi\sigma^{2}z\cosh\!\left(\frac{(F-F_{0})\Gamma}{2\xi}\right)\left[\frac{4P_{0}\xi}{\Gamma}+2z\sinh\!\left(\frac{(F-F_{0})\Gamma}{2\xi}\right)\right]}{\Gamma\left(\frac{4\xi^{4}}{\Gamma^{4}}+z^{2}\sigma^{4}\cosh^{2}\!\left(\frac{(F-F_{0})\Gamma}{2\xi}\right)\right)^{3/2}}\nonumber\\
&\!\!\!\!\times&\!\!\!\! \re^{(F-F_0)}\exp\!\left[-\frac{2\xi^{2}\sigma^{2}\left(\frac{2P_{0}\xi}{\Gamma}+z\sinh\!\left(\frac{(F-F_{0})\Gamma}{2\xi}\right)\right)^{2}}{\Gamma^{2}\left(\frac{4\xi^{4}}{\Gamma^{4}}+z^{2}\sigma^{4}\cosh^{2}\!\left(\frac{(F-F_{0})\Gamma}{2\xi}\right)\right)}\right].
\ea
The above integrals, obtained in the small-$\sigma$ regime, can be further simplified to extract general analytic results. First, one observes that 
\be{apprin}
z \cosh\left(\frac{(F-F_{0})\Gamma}{2\xi}\right)=\sqrt{\frac{4\xi\lambda}{3}}\re^{3(A+F_0)+3\Delta F}\cosh\left(\frac{\Delta F\,\Gamma}{2\xi}\right)
\ee
where $\Delta F\equiv F-F_0$ and $\Gamma/(2\xi)\equiv3+\epsilon^2>3$. Thus, once $\xi$ is fixed, $\re^{3\Delta F}\cosh\left(\frac{\Delta F\,\Gamma}{2\xi}\right)$ has a minimum at
\[
\Delta F_{\rm min}=\frac{1}{6+2\epsilon^2}\ln\pa{\frac{\epsilon^2}{6+\epsilon^2}}.
\]
Correspondingly, 
\be{apprinMIN}
z_{\rm min}=\sqrt{\frac{\xi\lambda}{3}}\,\re^{3(A+F_0)}\Bigg[\pa{\frac{\epsilon^2}{6+\epsilon^2}}^{\frac{6+\epsilon^2}{6+2\epsilon^2}}
+\pa{\frac{\epsilon^2}{6+\epsilon^2}}^{-\frac{\epsilon^2}{6+2\epsilon^2}}\Bigg]
\ee
and one can always choose $A$ properly to make it large enough. Thus, for small $\sigma$ and sufficiently large $A$, which is the relevant limit for calculating the spectrum, one finds at leading order 
\be{i0LO}
I_0={\widetilde{\mathcal{N}}}^2\sqrt{\frac{3\pi}{2\xi\lambda\sigma^2}}\re^{-3(A+F_0)}\Rightarrow {\widetilde{\mathcal{N}}}^2=\sqrt{\frac{2\xi\lambda\sigma^2}{3\pi}}\re^{3(A+F_0)}.
\ee
Using the normalization factor from (\ref{i0LO}) and the same approximations one also finds
\be{i1rsL0}
I_{1,r}=2P_0,\quad I_{1,s}=-2i
\ee
and the resulting, averaged spectrum is finally
\be{PS3}
\av{v_k^2}_A=c_k^{(0)}\!+2 s_k^{(1)}\!+2 r_k^{(1)}P_0
\ee
where the $P_0$-independent correction is modified by the Gaussian integral and the $P$-dependent part is essentially that which can be obtained with the standard semiclassical formalism. As in the GR case, the constant contribution is modified by the Gaussian distribution and one expects that higher order contributions in $P$ to $f(A;P,\bar P)$ can further modify the $P$-independent part of the spectrum. Let us note that such contributions are perturbatively small but not as small as the quantum-gravitational corrections which are usually calculated in such a context. 
\section{Conclusions}
Within the Wheeler-DeWitt description of the early Universe, we have investigated corrections to the primordial spectra arising from the quantum nature of the homogeneous inflaton-gravity system. These corrections are conceptually distinct from the quantum-gravitational corrections studied in a series of papers in recent years~\cite{AlberghiCasadioTronconi2006,KTV2013,KTV2014,KTV2015,KTV2016,KieferKramer2012,BiniEtAl2013,BrizuelaKieferKramer2016dS,BrizuelaKieferKramer2016SR,Calcagni2013}. Although they are still expected to be small, they may be significantly larger than the genuine quantum-gravitational contributions, which are suppressed by factors of order $\m^{-2}$ and are therefore extremely small. Moreover, the latter predominantly affect the lowest CMB multipoles, where observational uncertainties are large, particularly due to cosmic variance. In contrast, the quantum effects investigated in this paper, while being expected to mainly affect the largest scales, may also modify the spectrum at smaller scales and, at least within the simplified models considered here, can produce substantially larger corrections, depending on the wave-packet width $\sigma$. In particular, deviations at the percent level are possible. However, a reliable quantitative assessment of their observational relevance would require extending our analysis to more realistic inflationary models and performing a comparison with observational data.

The standard treatment of inflationary perturbations relies on the assumption that the homogeneous degrees of freedom involved in the inflationary dynamics, typically the inflaton field and the scale factor, behave classically and determine the evolution of the inhomogeneous perturbations, which are treated quantum mechanically. Ultimately, however, all degrees of freedom must obey the laws of quantum mechanics, including the homogeneous ones, and their quantum behavior should therefore be properly taken into account. The possibility explored here consists in considering the full quantum description of inflation provided by the Wheeler-DeWitt equation and a quantum superposition of solutions to its homogeneous sector forming a Gaussian wave packet. Although other choices may also be investigated, a Gaussian packet is usually regarded as the quantum wave function that most closely reproduces the classical evolution. Consequently, the resulting primordial spectra differ from the standard ones only through the quantum corrections in which we are interested (essentially associated with the width of the wave packet).

In general, one expects observables calculated within the full quantum description to be modified by the entanglement between the homogeneous degrees of freedom and the perturbations. Although this is generally the case, we have restricted our analysis to the entanglement generated by the so-called ``introduction of time.'' Heuristically, this effect may be understood by observing that the different trajectories contained in the wave packet of the homogeneous system are associated with different evolutions of the perturbations. The resulting perturbation spectra must therefore be obtained by ``averaging'' over these different trajectories, and the spectra derived in this way are indeed modified. Additional effects are expected because of the explicit dependence of the Mukhanov-Sasaki variable on the homogeneous inflaton and scale factor. Such effects are of order $\hbar^2$; however, their treatment is cumbersome and not unambiguously defined within the present framework, and we have therefore neglected them.

Finally, we have calculated the modified inflationary spectra for two simplified models: a minimally coupled inflaton with a constant potential and an induced-gravity inflationary model with a quartic potential. Both models admit a de Sitter attractor, but their classical trajectories approach this attractor with different velocities. Within the approximations adopted to obtain analytical expressions for the modified spectra, we have found that, in both cases, the leading-order, $P$-independent contributions [the constant parts in $f(A;P,\bar P)$] are modified. In the minimally coupled case, however, the calculation is less cumbersome and requires fewer approximations. As a by-product of our analysis, we have also found that, during the earliest stages of inflation, Gaussian wave packets may slightly deviate from the corresponding classical trajectories, which are recovered only at later times. This is an interesting result in its own right and deserves further investigation in future work.


\end{document}